\documentclass[12pt]{article}
\usepackage[english]{babel}
\usepackage{amsmath,amsthm}
\usepackage{amsfonts}
\date{}
\begin{document}
\def\thebibliography#1{\section*{REFERENCES\markboth
 {REFERENCES}{REFERENCES}}\list
 {[\arabic{enumi}]}{\settowidth\labelwidth{[#1]}\leftmargin\labelwidth
 \advance\leftmargin\labelsep
 \usecounter{enumi}}
 \def\newblock{\hskip .11em plus .33em minus -.07em}
 \sloppy
 \sfcode`\.=1000\relax}
\let\endthebibliography=\endlist

\hoffset = -1truecm \voffset = -2truecm


\title{{ QFT In Physical Systems: Condensed Matter to Life Sciences}
\author{A.N. Mitra \thanks{e.mail: ganmitra@nde.vsnl.net.in} \\ Dept of Physics, Univ of Delhi, Delhi-110007, India }}
\maketitle{}

\begin{abstract}
A panoramic view is given of the applications of QFT to diverse
areas of physics, beginning with the concepts of QED Vacuum as well
as of electron self-energy and renormalization. The related concepts
of the degenerate vacuum, spontaneous symmetry breaking (SSB)and
Nambu-Goldstone bosons, together with their applications to
condensed matter physics and quasi particles are outlined. The
possibilities of application  to life sciences are also examined.
Finally the alternative method of Path integrals for introducing QFT
is briefly indicated.
\end{abstract}

\section*{1  Introductory background}

Quantum Field Theory (QFT) has had a long history of evolution since
the time Dirac postulated his famous `Sea' of  negative energy
states, to the present day when efforts are seriously on for the
application of its methodology all the way to Brain Dynamics [1]. It
all started with the  single particle non-relativistic quantum
mechanics (QM)  formulation  by Heisenberg and Schroedinger who did
not have to face any consistency problem, until Dirac --and
Klein-Gordon before him--  introduced relativity into the picture.
Noting the failure of the (second order ) Klein-Gordon equation  to
produce a positive definite probability density, Dirac  attempted a
first order equation in matrix form (did he have the analogy to the
Maxwell equations in mind ?), and thus avoided the problem of
negative probability,  but only to find a second bottleneck-the
existence of negative energy states !  He eventually overcame this
problem  through his  novel theory of  the positron : a hole  in the
sea of negative energy states,  caused by the absorption of  enough
external energy ($ > 2 m c^2 $) by a (negative energy) electron to
get  lifted to a positive energy state, thus creating a partcle-hole
pair.   With the discovery (by C D Anderson) of  a cosmic ray
particle with precisely the properties of the "hole", he romped home
despite  Pauli's  determined  attack against his interpretation [2].
Yet Pauli must have gauged the true significance of Dirac's result,
as was to be evidenced by his own  successful quantization of  the
Klein-Gordon equation  [3] through a reinterpretion of the
non-positive definite probability density as an average charge
density of  both  types of  charges, positive and negative. Some
years later, Dyson [4] offered a clear perspective which covered
both the situations (Dirac sea of spin-half fermions as well as  the
Pauli-Weisskopf  average charge density of scalar bosons.) : For a
consistent theory which includes both Relativity and QM, one must
take  an infinite number of particles for mathematical
consistency-in other words  a field description with infinite
d.o.f.'s . In this sense  QFT  is the right generalization from the
QM  for a single particle ! Of course QFT had routinely been a part
of quantum theory since  the early thirties ,  but its true
significance  became clear after this conceptual input by Dyson [4];
see also Mitra [5].
\par
This paper is intended to serve as an elementary introduction to the
application of QFT methodology to `soft' condensed matter physics as
a preliminary to applications to life-science (with a proper role
for the environment)[1]. To that end we collect some essential
preliminaries in the next two sections on the methodology of QED (
the first non-trivial example of QFT), including definition of the
vacuum state (Sect 2), and ideas of self-energy and renormalization
(Sect 3). The concept of the degenerate vacuum is introduced next in
Sect 4, including those of Spontaneous Symmetry Breaking (SSB) and
long range  Nambu-Goldstone Bosons. The applications of QFT to
standard Condensed Matter Physics (CMP), with special reference to
phonons and magnons,  are outlined in Sect 5, while some additional
techniques relating to life sciences (which must now include the
role of environment as well as unidirectional flow of time ) are
taken up in Sect 6.Finally in Sect 7 the method of Path integrals is
outlined as an alternative means to access QFT, especially for its
role in incorporating the effect of environment for an open system.

\section*{2 QFT language: Vacuum State}

The QFT generalization of  single particle QM  is now routinely
considered as  the basic framework  for applications of quantum
theory to  diverse physical problems. The specific language employed
for the purpose is almost exclusively that of a harmonic oscillator
characterized by the  appearance of both the canonical variables
$\{x, p\}$ in a  quadratic  form in the energy operator , which in
certain units is  expressible as $ H = (p^2 + \omega^2 x^2)/2 $. And
because of the  symmetrical appearance of $\{x, p\}$, it is usually
convenient to  express the energy operator  in terms of their
standard complex combinations $\{ a, a^+\}$ which in  dimensionless
units read
\begin{equation}\label{2.1}
a = ( - i \omega x \hbar +  p ) / \sqrt { 2 \omega \hbar}; \quad
a^+ = ( + i \omega x\hbar+ p ) / \sqrt {2 \omega \hbar}
\end{equation}
$a$ and $a^+$ satisfy the commutation relations $[a, a^+]$ = 1,
while other commutators vanish. The energy operator, $H$ = $\hbar
\omega (a^+a + 1/2)$ incorporates the essential dynamics in terms of
the number operator $a^+a$ whose  integer eigenvalues $n$ are called
occupation numbers and the corresponding `states' (wave functions)
are called the eigenstates $| n >$ of $a^+a$.  By  virtue of the
equalities  $a | n
>$ = $\sqrt{n} | n-1 >$ and  $ a^+ | n >$ = $\sqrt{n+1} | n+1 >$,
$a$ and $a^+$ are called destruction / creation operators
respectively, since they reduce / increase the occupation number in
a given state $| n >$ by one unit each.  Hence by successive
applications of the $a*$ operators, the eigenstates $|n>$  are all
expressible in terms of the  ground state (`vacuum') $ | 0 >$  as
proportional to  ${a^+}^n | 0 >$, noting that $ a | 0>$ =0 . In this
language the successive eigenvalues of the energy operator are $E_n$
= $\omega \hbar ( n + 1/2) $. The QFT generalization for an infinite
collection of  harmonic oscillators  $a_k$, ${a^+}_k$ indexed by k
(another integer) ,  is now immediate : Since  $ [a_k, {a^+}_k']$ =
$\delta_{k,k'}$  incorporates all the commutation relations,  a
general  eigenstate  of this infinite collection of harmonic
oscillators $| \{ n_k\} >$ is compactly expressible as $\Pi _k
{{a^+}_k}^ {n_k} | 0 >$, where $| 0 >$ now represents the "master "
ground state, or Vacuum, of the infinite collection of harmonic
oscillators. [There is no problem  of `ordering' of these creation
operators since  they all commute].

This " master"  ground state -- or  simply Vacuum state -is a
central  theme around which  the entire  concept / methodology  of
QFT  devolves. [Rather paradoxically, the vacuum  concept  owes its
origin to Dirac's  Hole Theory not only for  fermions (for which it
was intended) but also for bosons through an obvious extension of
the ideas]. And in those days the vacuum was considered as a unique
ground state precisely of the type  described in the foregoing. The
earliest application of  the QFT formalism was in the area of
quantum electrodynamics -QED for short-the theory of interaction of
charged particles (electrons, free or bound) with the bosonic
electromagnetic (e.m.) field. This of course required a `vector'
formulation of  the  harmonic ocsillator field, a feature  already
present in the energy density ( $ {\bf E}^2 / 2 + {\bf H}^2 / 2$ )
of the Maxwell field. Its  Fourier analysis, apart from routine
technical details,   yields  creation and destruction operators of
the e.m. quanta together with  a  polarization index $\lambda$,
viz., $ a_{k\lambda}$, ${a^+}_{k\lambda}$,  which stand respectively
for the absorption / emission of a radiation quantum of momentum $k$
and polarization $\lambda$,  by  an electron.

\section*{3 QED: Self-Energy and Renormalization}

QED has had an intense two-decade old history (1930-50) of
vicissitudes with regard to experiment, wherein its spectacular
successes for lowest order processes like Compton scattering
(scattering of e.m. radiation  quanta by free electrons) and
Bremsstrahlung (scattering of electrons by  the coulomb field of an
atomic core, with the simultaneous emission of a quantum of
radiation) have co-existed with  total failures for processes
involving higher order e.m. effects, namely, prediction  of
infinities  which are totally incompatible with experiment.
Typically, a higher order e.m. process consists in  the emission and
re-absorption of a photon, grafted on an otherwise lowest order
process (such as Compton scattering above);  this  may be described
as a second order correction to the latter.  Now   since
energy-momentum conservation does not put any restriction on the
magnitude of the (virtual) photon momentum emitted and reabsorbed by
the electron,  the amplitude for the full process must be a sum of
all  sub-processes corresponding to all  possible momenta from zero
to infinity. And this sum is what gives rise to a (linearly)
divergent result on integration over k. This kind of
divergence-called the electron's $self-energy$ by virtue of the
capacity of the electron to emit and reabsorb a photon-- is typical
of the problems faced with QED processes whenever higher order
corrections to  a given  lowest order process  are attempted.
Perhaps these had best be ignored, as some physicists had argued
then, were it not for an unexpected development in the mid Forties
bearing on experiment: The microwave techniques developed during
World War II facilitated more accurate measurements of the energy
levels of atoms like hydrogen and helium than had been possible
hitherto. In particular it was found by Lamb and Retherford that the
$2p_{1/2}$ and $2s_{1/2}$ levels of hydrogen differ from the
original Dirac prediction of identical values (hitherto in excellent
agreement with experiment) by about 1040 megacycles ! This led to
intense discussions in which the infinities inherent in the second
order radiative corrections to atomic processes of the type
described above  figured prominently. Eventually at the Solvay
Conference of 1947 it was agreed that the idea of Renormalization
should be seriously considered for interpreting the electron's e.m.
self-energy ( defined above)  as  indistinguishable from its total
observed mass.  Now since the value of the electron's self energy
varies according to its environment,  it became logical--for
purposes of lumping this infinity with the mechanical mass of the
electron-- to identify some universal value which should be
independent of its environment; a natural candidate being the
self-energy of a $free$ electron ! Therefore the prescription became
clear : From the self energy of an electron bound in, say, a
hydrogen atom, subtract a universal part independent of its
environment, viz., that of a free electron. The resultant  quantity
should hopefully be finite  and hence measurable. The programme,
although attractive, had to be extremely elaborate so as to make the
result truly independent of any observer. This in turn needed a
fully covariant formulation  of QED, one  in which three stalwarts
(Tomanaga, Schwinger, Feynman) participated,  catalysed by a fourth
one (Dyson), and the entire programme took several years for the
essential features of the new theory to be fully implemented  In the
meantime a provisional estimate by Hans Bethe [6] indicated
excellent agreement with experiment. The rest  is history.

\section*{4 Degenerate Vacuum : SSB  and  NG-Bosons}

So far we have considered the interaction of the pure electron and
e.m. fields ( QED) whose  vacuum  is unique (non-degenerate) i.e., a
well-defined state of \emph{lowest} energy. In this background   the
concepts of Self -energy, and Renormalization discussed above  must
be regarded as second (and higher) order e.m. effects. Now the
uniqueness of the QED  vacuum  is the result of certain  standard
symmetries-translational, rotational, Lorentz-as well as  the
non-trivial symmetry of gauge invariance  which accounts for the
conservation of charge. Let us next consider a more general QFT (in
the background of other fields)  when one or more of such symmetries
are no longer satisfied. In such a generalized QFT scenario, the
Vacuum is no longer unique, rather \emph{degenerate}, characterized
by the phenomenon of   Spontaneous Symmetry Breaking (SSB) to the
accompaniment  of  certain \emph{massless}(as well as spinless !)
quanta called Nambu-Goldstone (NG) bosons (which necessarily have an
infinite range). The historical developments in this sector have
been aptly summarised by Weinberg [7]  in terms of the role of
symmetries in the development of the physics of the Twentieth
Century, from spacetime to internal symmetries. Thus `spontaneously
broken' symmetries are those that are not realized as symmetry
transformations of the physical states of the theory, and are always
associated with a degeneracy of vacuum states.  In particular, for
spontaneously broken continuous symmetries  there is a theorem
that,for each broken symmetry, the spectrum of physical particles
must contain one particle of zero mass and spin. [The theory must
not dictate which member is distinct, only that one of the members
is ].  Such particles-called Nambu-Goldstone bosons-- which
correspond to the spontaneously broken internal symmetry generators,
and are characterized by their quantum numbers-- were first
encountered by Nambu [8] in the context of BCS superconductivity, as
well as by Goldstone [9]  in a specific model; after this  two
general proofs of their existence were given by Goldstone, Salam and
Weinberg [10] within the framework of QFT.

These spinless bosons  transform nonlinearly (shift) under the
action of the generators of SSB, and can thus be excited out of the
( now asymmetric) vacuum by these generators. Thus, they can be
thought of as (collective) excitations of the field in the broken
symmetry directions in group space--and are massless if the
spontaneously broken symmetry is not also broken explicitly. If, on
the other hand, the symmetry is not exact, i.e., if in addition to
being spontaneously broken,  it is also explicitly broken, then the
Nambu-Goldstone bosons are not massless, although they typically
remain relatively light; they are then called pseudo-NG bosons.
Condensed matter physics abounds in  pseudo NG bosons, or  simply
pseudoparticles. Some typical examples are :
\par
i)   In fluids, the phonon is longitudinal ; it is the NG boson of
the spontaneously broken `Galilean' symmetry (wrt the transformation
$x'$ = $x - vt$). On the other hand in solids,  the NG bosons
(phonons) have both longitudinal and transverse modes, corresponding
to spontaneously broken Galilean, translational, and rotational
symmetries; but there is no simple one-to-one correspondence between
such modes and the broken symmetries. \\
ii)   In magnets, the original rotational symmetry (when no external
magnetic field is present)  is spontaneously broken such that the
magnetization points into a specific direction. The NG bosons  are
then called  magnons, i.e., quanta of  spin waves in which the local
magnetization direction oscillates. In addition if an external
magnetic field is present, the rotational symmetry is also
explicitly broken, in which case the magnons  also acquire a small
mass. \\
iii) As a third example--  this time from the field of elementary
particles-- the pions are the pseudo-NG bosons that result from the
spontaneous breakdown of the chiral  symmetries of QCD as a result
of  quark condensation (a typical  strong interaction effect). These
symmetries are further explicitly broken by the masses of the
quarks, so that the pions are not entirely massless, but their mass
is significantly smaller than typical hadron masses.

\section*{5 SSB in condensed matter physics}

Perhaps the most extensive use of  the concept of a degenerate
vacuum in QFT  has been  in the domain  of condensed matter physics
(CMP) where the facilities of experimentation are more readily
available than in the (more expensive) domain of particle physics.
This makes it possible to test the more intricate features of the
theory of vacuum structure in much greater detail than would be
feasible in any other comparative field. Indeed  in the CMP  sector,
there is a virtual goldmine of the corresponding NG bosons  --
phonons, magnons, plasmons, excitons, polarons, polaritons, to name
a few--, giving rise to a spate of discoveries like
superconductivity and superfluidity; Bose-Einstein condensation and
Josephson effect, all of which have been experimentally confirmed.
For purposes of illustration however we choose only two typical
NG-bosons -- phonons and magnons --. and briefly  indicate the steps
for their dynamical working as  a consequence of  SSB  in the
corresponding vacuum (including the Bogoliubov transformation).In
the following it will be instructive to quote selectively  from
Kittel's book [11] to illustrate the ideas involved.

\subsection*{5.1  Phonons as NG bosons}

Our first example of SSB in QFT -- the phonon  --   is the name for
the field  quanta of elastic excitations in a  crystal. In terms of
the  creation and annihilation operators ( $a_k$  , $a^+_k $ ) , SSB
generally causes  their mixing so as to violate  the standard
conservation of the corresponding QFT number operators $a_k^+a_k$.
Such  mixing in turn leads to certain  linear combinations
($\alpha_k$ and $\alpha^+_k$) emerging out of the original $a_k$,
$a^+_k $ pairs ( known as  the Bogoliubov transformation),  thus
defining new (conserved) number operators $\alpha^+_k\alpha_k $,
whose quanta are now the NG bosons. However, in the simplest case of
elastic lattice (discrete) vibrations from crystals  one already
obtains a diagonalized Hamiltonian (even before a Bogoliubov
transformation!) in the standard form [12]  :
$$ H = \sum  \omega_k ( n_k  + 1/2 ) ;  \quad \omega_k = \sqrt{2(1-\cos k)} ; \quad  n_k = a_k^+ a_k  $$
The number of phonons $(n_k)$, in certain units,  in the state k is
a non-negative integer.  As a less trivial example, phonons can
arise in a condensed Bose gas of weakly interacting particles
through a similar treatment, but now there also appear terms of the
type $a_k$, $a_{-k}$ and $a_k^+$, $a_{-k}^+$ in the Hamiltonian, and
one now needs for its diagonalization  a Bogoliubov transformation
[12] :
$$ \alpha_k = u_k a_k - v_k a_{-k}^+ ; \quad \alpha_k^+ = u_k a_k^+ - v_k a_{-k} $$
where $(u_k, v_k)$ are real functions of k (momentum). The
corresponding dispersion relation (behaviour of $\omega_k$ as a
function of k) is a straight line for small k (acoustic mode),  but
has a pronounced dip  at  larger k ; this dip  has a built-in
mechanism for understanding the phenomenon of superfluidity in a
condensed Bose - Einstein gas in liquid helium 4 near absolute zero
of temperature (see [12] for details).

\subsection*{5.2  Magnons as NG bosons}

A second example  of  QFT techniques  concerns the rotational
dynamics of spin  [13]  which  encompasses  the all-important
concepts of spin magnetic moment and the associated magnetic field
(internal and external). With  spin d.o.f.'s  playing the main role,
the low-lying energy states of such  systems coupled by exchange
interactions are wave-like,  giving rise to `spin waves' . The
quantization  of their energy then gives  rise to  quanta called
magnons. The  basic  dynamics of such spin systems   with nearest
neighbour interactions in the presence of an external magnetic field
$ H_0$ (along the z-axis)  is given by the Heisenberg interaction
Hamiltonian applicable to a ferromagnet :
$$ H  =  -J  \sum {\bf S}_j \cdot {\bf S}_{j+ \delta } - g \mu_B H_0 \sum S_{j z}  $$
where $J > 0$ is the exchange integral ; $\delta$ defines the range
of nearest neighbour interactions; $g \mu_B$  is the magnetic
moment; and ${\bf S_j}$ is the spin angular momentum of the atom at
location j. The  (x,y,z) components of each  ${\bf S_j}$ are
connected by the (j-independent) condition [13]
$$ {\bf S_j}\cdot {\bf S_j} =S(S+1)$$

The crucial step needed to bring in the standard creation and
destruction operators  of QFT is now the Holstein-Primakoff
transformation [13] :
\begin{eqnarray}\label{5.1}
S_j^+  &=& S_{jx} +  iS_{jy} = \sqrt{2S} \sqrt{( 1 - a_j^+ a_j /
2S)} a_j \\ \nonumber
 S_j^- &=& S_{jx}  -  iS_{jy} = \sqrt{2S}
\sqrt{( 1 - a_j^+ a_j / 2S)} a_j
\end{eqnarray}
The number operators $a_j^+a_j$'s in turn are connected to the
$S_{jz}$'s  by the relation $S_{jz}$ = $ S - a_j^+ a_j$. The
transformation from the atomic $(a_j)$ to the magnon $(b_k)$
variables is now achieved by the usual Fourier transformation
connecting the two sets [13],  whence the latter can be shown to
satisfy the standard commutation  relations $ [b_k, b_k'^+] $ = $
\delta_{kk'}$. The total spin operator $\mathcal{S_z}$ is now
expressed by
$$ \mathcal{S_z} = NS - \sum b_k^+ b_k $$
where N is the number of atoms each of spin S.  With the help of
these results,  the magnon Hamiltonian  becomes a function of the
${b_k, b_k^+}$ variables which can be expanded perturbatively in
powers of the scalar ${\bf k}\dot {bf \delta}^2$.  This quantity
which  reduces to $(ka)^2$ for  several  standard lattices with a
lattice constant a,  is usually small, thus justifying a
perturbation treatment. The diagonalization of the Hamiltonian --
without a Bogoliubov transformation --  then  yields a dispersion
relation of the form [13] :
$$ \omega_k = g\mu_B H_0 + 2JS (ka)^2 $$
This result agrees with  the  corresponding data on magnon
dispersion relations in magnetite as determined from  inelastic
neutron scattering [13]. A similar treatment goes through for an
antiferromagnet as well, but  now the effective d.o.f.'s get doubled
since the spin structure of the crystal get divided into two sets of
interpenetrating sublattices $a$ and $b$ with the property that all
nearest neighbours of an atom on $a$ lie on $b$ and vice versa [13].
As a result we now have two sets of  QFT operators $a_j, a_j^+$ and
$ b_j, b_j^+$ which therefore require a Bogoliubov transformation
for the diagonalization of the corresponding Hamiltonian. The
dispersion relation for $\omega_k$ is now more complicated, but
shows a linear behaviour for a simple cubic lattice for small $ka$.

\section*{6 QFT Application to life sciences}

We are now in a position to extend the QFT formulation to soft CMP for possible
applications to biological systems. There are now two additional features to be
taken into account: i) effect of environment; and ii)  unidirectional flow of time
(amounting to a dissipative dynamics). To implement these features we borrow a model
of brain dynamics due to Freeman and Vitiello [14], which is an extension--
to dissipative dynamics-- of a many-body model proposed by Umezawa et al [15, 16] which states[14]: \\

 ``in any material in CMP any particular information is carried by certain ordered patterns
 maintained by certain long range correlations mediated by massless quanta. It looked to me
 (Umezawa) that this is the only way to memorize some information; memory is a particular pattern
 of order supported by long range correlations." \\

The long range correlations dynamically generated by SSB via
NG-bosons (zero mass modes spanning the whole system) are thus
crucial to the`memory' mechanism envisaged by Umezawa. And the
coherent condensation of NG bosons in the ground state (vacuum) of
the system gives it the appearance of an `ordered' state.  A measure
of the degree of ordering (or coherence) is provided by the vacuum
density of the NG bosons. This measure is also called the order
parameter which is a classical field "labelling" the observed order
pattern [17]. The recall of ordered information occurs under a
stimulus able to excite the density wave quanta (DWQ)out of the
ground state (vacuum). Such a stimulus is called "similar" to the
one responsible for the memory recording.

Unfortunately this many-body model by itself cannot explain the
observed co-existence of amplitude modulation (AM) patterns and
their irreversible time evolution [ 17]. For, any subsequent
stimulus `cancels' the previous recorded memory by renewing the SSB
process, thus overprinting the new memory over the previous one--a
momory capacity problem [17]. According to del Giudice et al [17],
as well as Vitiello [14], the many-body model [15] did not consider
an \emph{open} system in permanent interaction with the environment,
thus  missing the key ingredient of \emph{Dissipation}.
\par
Now Dissipation is a relatively new idea whose insertion into the
QFT framework involves borrowing from both the picture of the `Dirac
sea'( playing the role of environment), as well as Feynman's novel
concept of "Zig-zagging in time" [18], a feat which Vitiello et al
[19] performed by \emph{doubling} the system's d.o.f.'s as follows

\subsection*{6.1 Environ as time-reversed image of system}

External stimulus leads to SSB which in turn causes dynamical
generation of density wave quanta $A_k$. A crucial step is now to
double  the number of d.o.f.'s to $\{A_k, {\tilde A}_k\}$, where
${\tilde A}_k$ is the time reversed mirror image. [ The analogy is
to the `method of images' in electricity to ensure
`equipotentiality' of the surface in front of a charge, the surface
now playing the role of `environment'!]. Next define the
corresponding number operators $N_{Ak}$ = $A_k^+ A_k$, together with
a similar expression for  $N_{{\tilde A}k}$. Then the energy flux
balance is
\begin{equation}\label{6.1}
E_0 = E_{syst} - E_{env} = \sum \hbar \Omega_k ( N_{Ak}- N_{{\tilde
A}k}) =0
\end{equation}
Note the similarity of the operators $\{A_k, {\tilde A}_k\}$ to the
electron and `hole' d.o.f.'s in the Dirac sea. To pursue the analogy
further, in CMP, the Fermi sea has $E < E_F$ for hole states, and $E
> E_F$ for electron states. Again in nuclear physics, the ground
state corresponds to the vacuum, while the excited states consist of
particle-hole pairs.
\par
But the environment is a more complex system in which
\emph{dissipation} is a key feature manifesting through some sort of
`damping' in the energy operator arising from the two sets $A_k$ and
${\tilde A}_k$ of quasi massless operators, while their commutation
relations retain their  standard structures. The damping effect now
shows up through the part $H_I$ of the total energy operator $H$ =
$H_0 + H_I$,  while the `real part' $H_0$ is  symmetrical between
the two sets:
\begin{eqnarray}\label{6.2}
H_0  &=& \sum_k \hbar \Omega_k [A_k^+A_k - {\tilde A}_k^+{\tilde A}_k] \nonumber \\
H_I  &=& i \sum_k \hbar \Gamma_k [ A_k^+{\tilde A}_k^+ - A_k{\tilde
A}_k]
\end{eqnarray}
The two parameters $\Omega_k$ and $\Gamma_k$ can now be recognized
as the real and imaginary parts of the frequency variable relating
to the `k-mode'.  Note that the `damping' parameter $\Gamma_k$ does
not exist in standard QFT , while the energy flux balance
$$ \sum_k \hbar\Omega_k ( N_{Ak}- N_{{\tilde A}k}) = 0 $$
where the N's are non-negative integers, is satisfied in the usual
manner. The `vacuum' is now made up of two sets of zero occupation
numbers, so that states can be built up by a straightforward
doubling of the techniques of Sect 2. Further, in view of the
commutation relation,
$$ [H_0, H_I] = 0 $$
the combinations $( N_{Ak} -  N_{{\tilde A}k})$ are constants of
motion for any $k$, so that $H_0$ remains bounded from below if
bounded at some initial time $t_0$. For other details see ref [19].

The new theory produces several agreements with observations which
the many-body theory of Umezawa et al [15] could not explain. These
include the QFT dissipative dynamics;  (quasi-)non-interfering
degenerate vacua; AM pattern textures (phase transitions among
them); AM patterns sequencing; and huge memory capacity.

\section*{7 An alternative formulation : Path Integral Method}

For the sake of completeness, we shall outline an alternative
formulation for the introduction of QFT language in physics, viz.,
Feynman's Path Integral method [20] (a book with no references !)
for a novel yet holistic  understanding of quantum mechanics
(without using the language of operators). In its simplest form, it
is a sum over all possible \emph{classical} trajectories, each with
a definite \emph{phase}which is  proportional to $\exp{i \int dt
L(x,t) /\hbar}$,where $\int dt L(x,t)$ is the classical ``action"
(time-integral over the Lagrangian) over a given path, so that the
total amplitude is a sum over all such contributions [20]:
\begin{equation}\label{7.1}
 \sum_{all paths from a to b} const  \exp{i\int dt L(x, t) /\hbar}
 \end{equation}
where the constant is subject to (later) adjustment to give the
correct normalization. Before proceeding further we pause to make
some preliminary comments on the significance of this fundamental
expression, especially its correspondence with the precise quantum
mechanical amplitude it is supposed to represent. Historically, this
form for the phase factor  was first conjectured by Dirac in 1931,
but not pursued further, until Feynman two decades later [21] gave a
more precise interpretation and developed the ideas  systematically
to show its equivalence to conventional quantum mechanics, together
with the rules of evaluation of the integrals associated with the
``master expression" (5) [20]. To describe its structure in a
nutshell, it represents the {\it complete  amplitude} for the
propagation of a particle from an initial space-time point $a$ to a
final space-time point $b$, (also called the Feynman propagator).In
the conventional quantum description, this quantity is the result of
a long-drawn calculation (using standard techniques)to arrive at
this elaborate structure. In this alternative description,on the
other hand, this ``master expression" is the {\it starting point}
from which it is possible to arrive at, say, the Schroedinger
equation by working backwards [20],and in so doing establishing a
consistency with the latter.
\par
The principal strategy for the evaluation of (5) is first to define
in a systematic way the $measure$ of the multiple integrations
involved therein, (all of which are amenable to successive gaussian
integrations), and then carry out the integrations through a
suitable `discretization' procedure, together with a final limiting
process--all in a highly pragmatic (no rigour !) manner.
 An important advantage of this ``Path Integral" method whose input is
 the classical action $S$ = $\int dt L(x, t)$ is that all quantum mechanical
 phases are already built-in within entirely \emph{classical } premises. And any
 quantum mechanical amplitude of physical interest can be derived from this form
 merely with the help of suitably-defined functional derivatives. In the general
 case when the quantum system consists of a mixture of states (not a pure one), the
 corresponding amplitude is called a {\it density matrix}.
 \par
 A second advantage of the Path Integral method is that it is trivially adaptable to
 the techniques of statistical mechanics merely with the
 replacement of $ i \int_0^t dt L(x, t)/\hbar$ by $-\int_0^{\beta\hbar} d\tau H(x,\tau)$,
 where the new variable $\tau$ = $-i t$ is purely imaginary (called the Matsubara time) and
 has a maximum value $\beta\hbar$ which is related to  the temperature $T$ by $kT$ = $1\\beta$ .
 [ Note that this has also necessitated a simultaneous replacement of the Lagrangian $L$ by the
 Hamiltonian $H$ ].  It is remarkable that with such a simple change from a real to imaginary `time'
 the scope of the Path Integral technique has got vastly extended to be applicable to the entire field
 of statistical mechanics, with almost no  extra charge. A more familiar name for the `density matrix'
 for the statistical system is the ``Partition function", which goes to show that latter is calculable
 in terms of the former, simply by extending the scope of the time variable to imaginary values
 (Matsubara time).

 \subsection*{7.1 Application to ``open" Systems}

Still another advantage of the Path Integral method is that with
little extra input, it can be extended to calculate the behavior of
a system of interest, even when it is coupled to other external
quantum systems, in terms of its own variables only. The necessary
formalism is due to Feynman and Vernon [22] who have shown that the
effect of the external systems in such a formalism can always be
included in a general class of functionals -- termed influence
functionals-- of the coordinates of the given system only. To that
end it is first necessary to extend the action functional for the
path integral to include the variables of the external system
(environment)as well. Such variables represent, e.g., the effect of
classical forces, linear dissipative systems at finite temperatures,
as well as combinations of them. The Feynman-Vernon method is
particularly applicable to linear systems composed of combinations
of harmonic oscillators, loss being introduced by continuous
distributions of oscillators. The resultant quantity may be called
the generalized density matrix for the combined system, which, after
integration over the external variables is called the reduced
density matrix, which principal components are the ``Influence
functionals".  The form of the latter is particularly simple for
linear dissipative systems, namely, they have the same form as
obtainable in terms of their classical response functions.





\end{document}